\documentclass[aps,prb,twocolumn,showpacs,superscriptaddress,amsmath,amssymb]{revtex4-1}

\usepackage{graphicx}  
\usepackage{bm}
\usepackage{xcolor}
\usepackage{gensymb}                  
\usepackage[unicode=true,pdfusetitle,
 bookmarks=true,bookmarksnumbered=false,bookmarksopen=false,
 breaklinks=false,pdfborder={0 0 1},backref=false,colorlinks=true]
 {hyperref}
\hypersetup{
 urlcolor=blue,linkcolor=blue,citecolor=blue,filecolor=blue}
 
\begin{document}

\title{Probing the role of carbon solubility in transition metal catalyzing Single-Walled Carbon Nanotubes growth}
\author{J.M. Aguiar-Hualde}
\affiliation{Laboratoire d'Etude des Microstructures, ONERA-CNRS, BP 72, 92322 Ch\^atillon Cedex, France}
\author{Y. Magnin}
\affiliation{Centre Interdisciplinaire de Nanoscience de Marseille, Aix-Marseille University and CNRS, Campus de Luminy, Case 913, F-13288, Marseille, France.}
\author{H. Amara}
\affiliation{Laboratoire d'Etude des Microstructures, ONERA-CNRS, BP 72, 92322 Ch\^atillon Cedex, France}
%
%
\author{C. Bichara}
\affiliation{Centre Interdisciplinaire de Nanoscience de Marseille, Aix-Marseille University and CNRS, Campus de Luminy, Case 913, F-13288, Marseille, France.}
\date{\today}

\begin{abstract}
Metal catalysts supporting the growth of Single Wall Carbon Nanotubes display different carbon solubilities and chemical reactivities. In order to specifically assess the role of carbon solubility, we take advantage of the physical transparency of a tight binding model established for Ni-C alloys, to develop metal carbon models  where all properties, except carbon solubility, are similar. These models are used to analyze carbon incorporation mechanisms,  modifications of metal / carbon wall interfacial properties induced thereby, and the associated nanotube growth mechanisms. Fine tuning carbon solubility is shown to be essential to support sustainable growth, preventing growth termination by either nanoparticle encapsulation or detachment. 
\end{abstract}

\maketitle

\section{Introduction}

Carbon nanostructures (graphene or nanotube) can be grown by chemical vapor deposition (CVD) techniques on various transition metals~\cite{Jourdain2013, Batzill2012} and alloys ~\cite{He2010, An2016, Dai2011}. In this process, a carbon bearing molecular precursor is decomposed at the surface of a catalyst to deliver carbon and feed the carbon network~\cite{Journet2012,Jourdain2013,Cabrero2016}. In case of single-wall carbon nanotubes (SWNTs), many applications require to control their synthesis at the nanoscale level to take advantage of their chirality-dependent properties. In this context, controlling the growth requires a detailed understanding of the role of the catalyst. However, the complex behavior of catalyst nanoparticles (NP)  exposed to reactive carbon makes it difficult to grow tubes with defined chirality. 

Efficient catalysts to grow SWNTs lie in the 1-10 nm diameter range and are reported to display non-zero, but limited carbon solubility in the bulk solid~\cite{Deck2006}. This is sometimes illustrated in the form of a so-called "volcano plot" showing that popular catalysts  (Fe, Co, Ni) display a large enough affinity for carbon, to adsorb carbon precursor molecules in the surface layers and possibly dissolve some carbon inside the NP. Indeed, this carbon content favors the dewetting of the nanoparticle with respect to the $sp^{2}$ carbon wall, a necessary property to limit catalyst encapsulation and deactivation~\cite{Diarra2012, He2015, He2017}. Despite such progresses, a notable lack of understanding of the roles played by carbon solubility in the catalyst during the SWNT growth process plagues our ability to optimize synthesis conditions and ultimately control  the structure of the tubes.

In transition metals, C occupies interstitial sites due to its relatively small atomic size~\cite{Toth1971, Cottrell1995,Siegel2003} leading to narrow domains of stability of solid solutions. The few experimental data found in the literature are devoted to bulk systems and report large discrepancies~\cite{Mclellan1969,Arnoult1972,Lopez2004,Portnoi2010}. From a theoretical point of view, Hu \textit{et al.}~\cite{Hu2015} employed first-principles calculations to study the solubility and mobility of B, C, and N interstitials in various transition metals, including Gibbs energy contributions at finite temperature. In the context of catalysis, subsurface interstitial sites have been identified as the most favorable for carbon incorporation on flat surfaces of different transition metal-carbon systems~\cite{Klinke1998, Teschner2008, Yazyev2008}. Besides, melting temperatures and carbon solubility limit have been investigated in case of Fe and Ni NPs used as catalysts to grow SWNTs~\cite{Ding2004b,Harutyunyan2008, Engelmann2014}. Recently, carbon rich phase diagrams of nickel-carbon nanoparticles have been calculated for system sizes up to about 3 nm, highlighting how C atoms can diffuse inside surface layers and induce a partial or complete melting~\cite{Magnin2015}. However, the case of such NPs in contact with tubes, which are much more complex, were not addressed and much remains to be done.\\

The aim of the present paper is to investigate to which extent growth mechanisms of SWNTs depend on the catalyst carbon solubility, and to check whether modifying this carbon solubility could lead to improved CVD synthesis in terms of yield and selectivity. Metal catalysts usually display different carbon solubilities and chemical reactivities. In order to specifically assess the role of carbon solubility, we take advantage of the physical transparency of a tight binding model established for Ni-C alloys~\cite{Amara2009}, to develop metal carbon models  where all properties, except carbon solubility, are similar.  By computing  carbon adsorption isotherms on various catalytic NPs, we first characterize their physical and chemical states in presence of carbon. Since wetting properties plays a major role in the growth mechanisms of SWNTs, we then directly simulate the structure of carbon enriched NPs on a graphene layer at different compositions. In the last part, we show how carbon solubility acts on the interplay between two mechanisms in competition : NP dewetting \textit{vs} wall growth.

\section{Methods}

To describe the Ni-C system, we have developed a model based on the tight-binding (TB) approximation which provides an efficient tool to calculate carbon and transition-metal interactions~\cite{Amara2009}. Only $s$, $p$ electrons of C and $d$ electrons of Ni are taken into account. Local densities of electronic states are calculated using a recursion method, where the first four continued fraction coefficients only are calculated exactly. Applications of this model to the catalytic growth of graphene and carbon tubes have already been presented elsewhere~\cite{Amara2008,Diarra2012,Weatherup2014}, showing the relevance and robustness of this model. A further advantage of our TB model is that it can be fairly easily generalized to other metal-carbon systems since we know qualitatively how the different parameters (transfer integrals, atomic energy levels, etc...) vary with the nature of the metallic element. We therefore take advantage of the physical transparency of the model to identify the parameters controlling carbon solubility. \\

To understand  carbon-metal interactions, it is convenient to study the electronic structure of metal carbides in a simple, and sometimes not physically observed, NaCl structure. Although they do not always exist for some transition metals, they can be studied on the basis of very robust first-principles calculations~\cite{Schwarz1977,Haglund1991}. As reported previously, the shape of the hybridized band does not change too much when varying the element along the transition metals series~\cite{Amara2009}. The ability of transition metals to combine strongly with carbon stems from the strong covalent bonds formed between the $p$ states of carbon atoms and the valence $d$ states of the metal ones~\cite{Schwarz1977,Tan1997}. As a result, by simply tuning the energy difference between the carbon atomic $p$ ($\epsilon_{p}^{C}$) and the metal $d$ ($\epsilon_{d}^{M}$) levels, we can change the M-C interaction and hence the carbon solubility in the metal. This is done by keeping all other parameters fixed, meaning that all properties such as melting point, lattice parameters, elastic constants and energy of pure metal and pure C are untouched. 

We thus probe only the effect of C solubility by changing the position of the atomic $d$ level which obviously varies with the nature of the element considered. Indeed, $\epsilon_{d}^{M}$ decreases when increasing the number of electrons along a transition series (about 1 eV per element)~\cite{Ducastelle1991,Pettifor1995, Tan1997}. Since this level is adjustable to some extent, it is useful to see how it is related to the carbon solubility.
\begin{figure}[htbp!]
\includegraphics[width=0.90\linewidth]{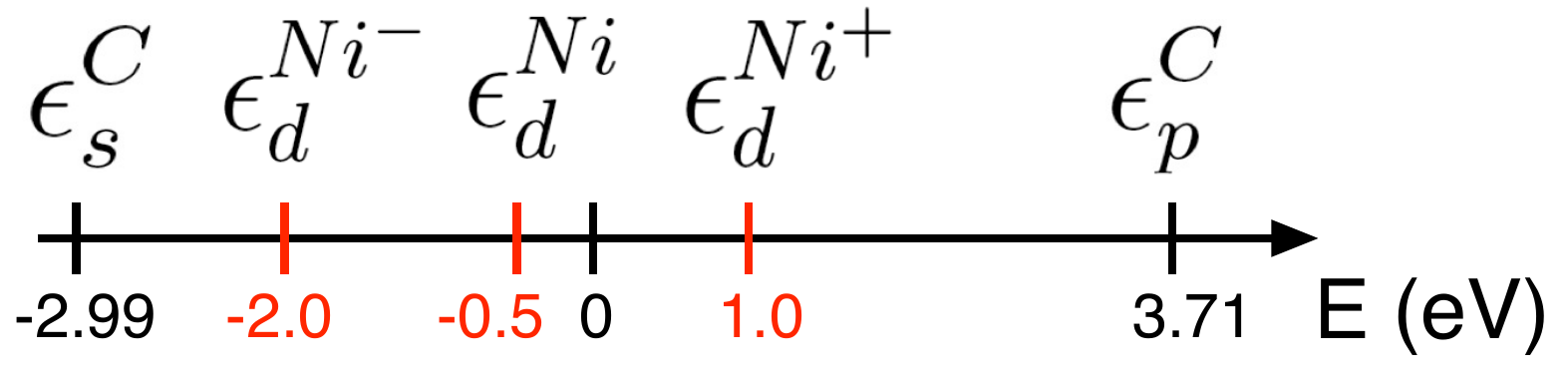}
\centering
\caption{Schematic representation of the atomic levels used in our tight-binding model.}
\label{Figure_1}
\end{figure}
To quantify this effect, let us consider the heat of solution $\Delta H_{sol}$ of a C interstitial atom in a crystalline transition metal (M), calculated according to the formula : 
\begin{equation}
\Delta H_{sol} = E_{M+C} - E_{M}  - E_{C}, 
\end{equation}
where $E_{M+C}$ is the total energy of the Metal+ interstitial C system, $E_{M}$ is the energy of the metal without C, and $E_{C}$ is the energy per C atom in graphene. In the FCC lattice, the most likely location for C in octahedral interstitial sites, which is confirmed by DFT calculations~\cite{Siegel2003, Cottrell1995}. For C, we assume the same atomic energy levels ($\epsilon_{s}^{C}$ = -2.99 eV and $\epsilon_{p}^{C}$ = 3.71 eV) as proposed by Xu \textit{et al.}~\cite{Xu1992} and modify the $pd$ hybridization through the variation of the metal $d$ level. All results are presented in Figure~\ref{Figure_1} and Table~\ref{tab: Delta Hs}. The $p-d$ energy difference ($\epsilon_{p}^{C}-\epsilon_{d}^{M}$) for Ni in our model is equal to 4.21 eV. This results in a heat of solution of one C in Ni  $\Delta H_{sol}=+$0.47 eV, in agreement with the +0.43 eV value found in the literature~\cite{Siegel2003, Hu2015}. To decrease C solubility, the contribution of the $pd$ hybridization has to be reduced. This can be done by shifting the relative position of the $p$ (C) and $d$ (M) atomic levels by 1.50 eV ($\epsilon_{p}^{C}-\epsilon_{d}^{M}=$4.21 eV) leading to a $\Delta H_{sol}$ around +0.87 eV. We also consider another case with a higher solubility where $\epsilon_{p}^{C}-\epsilon_{d}^{M}=$2.71 eV, yielding $\Delta H_{sol}=+$0.33 eV. For the sake of simplicity, let us denote Ni$^{-}$, Ni and Ni$^{+}$, metals with increasing C solubility respectively. At this stage, we have at our disposal a model based on the tight-binding approximation to describe transition metal-carbides with three different carbon solubilities. 
\begin{table}[h]
\small
\caption{Values of $\Delta H_{sol}$ for three different solubilities. }
\label{tab: Delta Hs}
  \begin{tabular*}{0.5\textwidth}{@{\extracolsep{\fill}}lll}
    \hline
 & $\epsilon_{d}^{M}$ (eV) & $\Delta H_{sol}$ (eV/at) \\
    \hline
Ni$^{+}$(high solubility) &+1.00 &+0.33 \\
Ni & -0.50 & +0.47 \\
Ni$^{-}($low solubility) & -2.00 & +0.87 \\
    \hline
  \end{tabular*}
\end{table}

This energy model is implemented in a Monte Carlo (MC) code using either canonical or Grand Canonical (GC) algorithms with fixed volume, temperature (T), number of metal atoms and C chemical potential ($\mu_{C}$)~\cite{Frenkel2002}. The GC algorithm used consists in a series of Monte Carlo cycles. Each cycle randomly alternates displacement moves for the metal and C atoms, attempts to incorporate C in a previously defined active zone and attempts to remove existing C atoms. 

\section{Carbon solubility in nanoparticles}

Both experimental and theoretical approaches point at the importance of the carbon fraction dissolved in the catalyst during the CVD synthesis of SWNTs~\cite{Diarra2012, He2015}. It is thus relevant to evaluate its  influence on the chemical and physical states of different NPs, by modifying the carbon solubility in the catalyst, using our models for Ni$^{-}$, Ni and Ni$^{+}$. 

\begin{figure}[htbp!]
\includegraphics[width=1.0\linewidth]{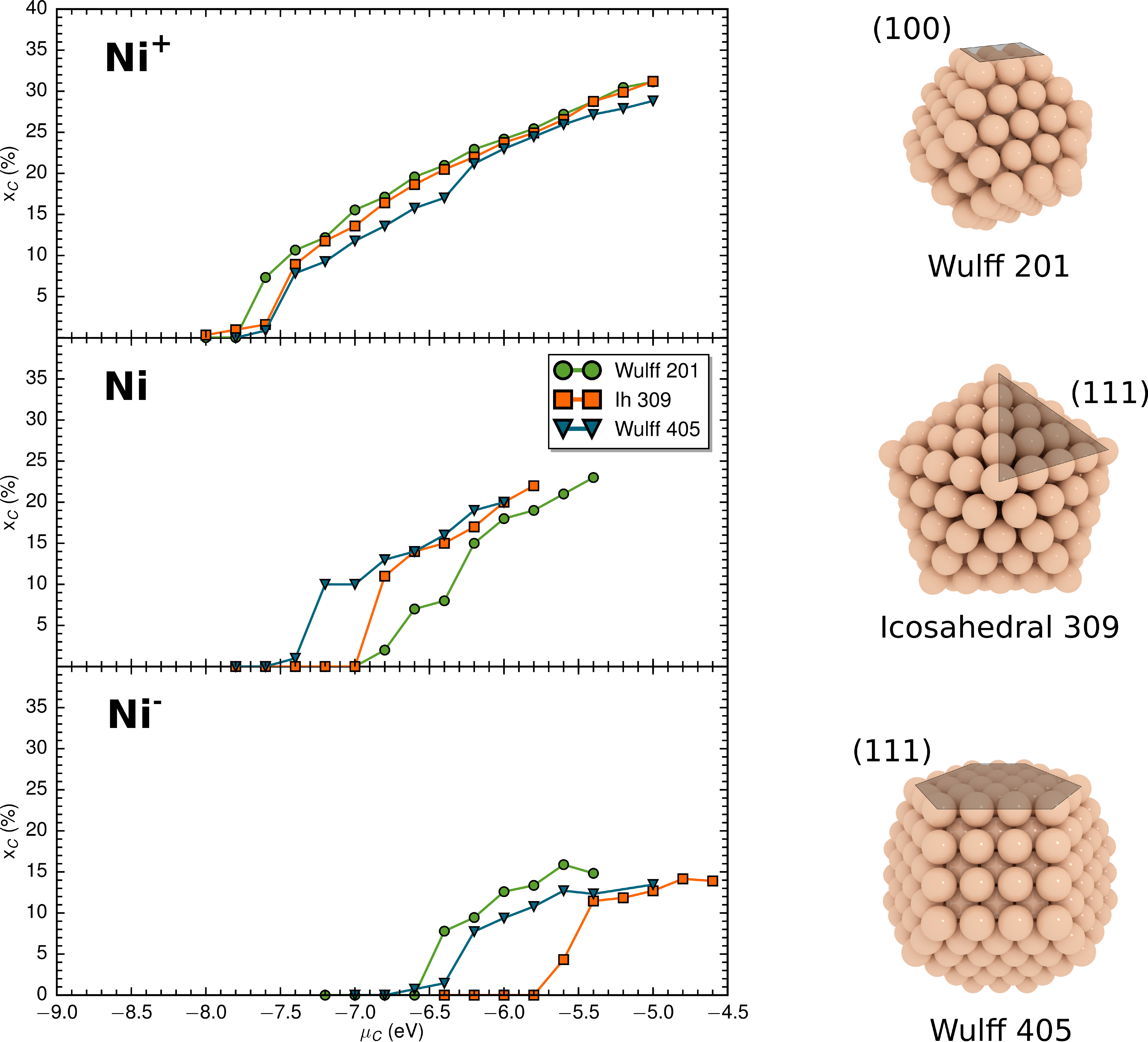}
\caption{(Left) Carbon adsorption isotherms calculated at $T = 1000 K$ on the basis of GCMC simulations for NPs (with 201, 309 and 405 atoms) with high (top) and low (bottom) carbon solubility. (Right) Schematic representation of Wulff and Ih structures where the different facets are presented. The NP with 309 atoms is Ih when pure, the others are FCC, with their Wulff equilibrium shape.}
\label{Figure_2}
\end{figure}

GCMC calculations presented here compare adsorption isotherms of carbon on NPs with different solubilities, sizes and structures. Initially Icosahedral (Ih) with 309 atoms, and FCC, Wulff shaped clusters with 201 and 405 atoms are considered (see Fig.~\ref{Figure_2}). These clusters have sizes around 1.8-2.5 nm, close to those used experimentally to grow tubes. Once the equilibrium is reached during GCMC simulations, we record the fraction of C atoms absorbed inside the cluster ($x_{C}$) at a given $\mu_{C}$ and $T$ to calculate the carbon absorption isotherms for different particle sizes. Some of the already published data are presented here for the sake of completeness~\cite{Magnin2015,He2015}. 
A number of conclusions can be drawn from the inspection of these adsorption isotherms at 1000 K presented in Fig.~\ref{Figure_2}. Beyond a certain $\mu_{C}$ threshold, the carbon fraction gradually increases until the carbon solubility limit is reached. This limit corresponds to the carbon fraction beyond which carbon atoms begin to appear on the surface of the NPs. As expected, the maximum solubilities of carbon in low-carbon-solubility metal NPs (Ni$^{-}$) and high-carbon-solubility (Ni$^{+}$) are $\sim15\%$ and 30\%, respectively. This is reasonable compared to the carbon solubility in normal Ni NPs which lies around 25\%. The threshold for the incorporation of the first C atom is also closely related to the nature of the catalyst. It is shifted towards higher (less negative) values, when C solubility in the metal decreases. This is not surprising since incorporating a C atom in a system that displays a lower solubility is more difficult, hence requires higher chemical potential. In addition, for Ni and Ni$^{-}$ cases, the relative position of the isotherm of the icosahedral NP tends to shift to larger $\mu_{C}$ values as compared to the FCC ones. The maximum difference between the Ih and FCC adsorption thresholds ($\Delta\mu_{C}$) is obtained for Ni$^{-}$, in which case the difference ($\Delta\mu_{C}\sim1$eV) is quite significant. Indeed, assuming an ideal behavior of the gas phase, the pressure is related to $\mu_{C}$ by: $\triangle \mu_{C} = k_{B}T \ln(p_{1}/p_{2})$. Accordingly, a value of $\triangle \mu_{C}$ around 1.0 eV per atom at 1000 K corresponds to $\sim5$ orders of magnitude pressure difference for Ih NP. This is clearly an effect of the structure of the NP. Ih NPs display only (111) facets while Wulff shaped NPs also display open (100) facets which are the most favorable surface adsorption sites for C~\cite{Magnin2015}. As a result, icosahedral NPs, where such sites are not present, require larger $\mu_{C}$ to stabilize adsorbed carbon. Moreover, this shift of the Ih NPs isotherms is clearly more pronounced for Ni$^{-}$ NPs, while it disappears for NPs with high solubility.  \\

To investigate the atomic structures of the NPs, we need to define a local order parameter $\bar{S}$. This can be done using the orientational order parameter introduced by Steinhardt \textit{et al}~\cite{Steinhardt1983} to discriminate between crystalline and disordered (liquid or amorphous) environments for each atom. Then, by averaging over all metallic atoms of the NP, it is possible to assign a global degree of crystallinity $\bar{S}$. Using this approach, the calculated phase diagrams of nickel-carbon nanoparticles have shown the presence of a large  crystalline core / liquid shell domain, instead of the liquid/solid two-phase domain characteristic of the bulk~\cite{Magnin2015}. 

\begin{figure}[htbp!]
\includegraphics[width=0.90\linewidth]{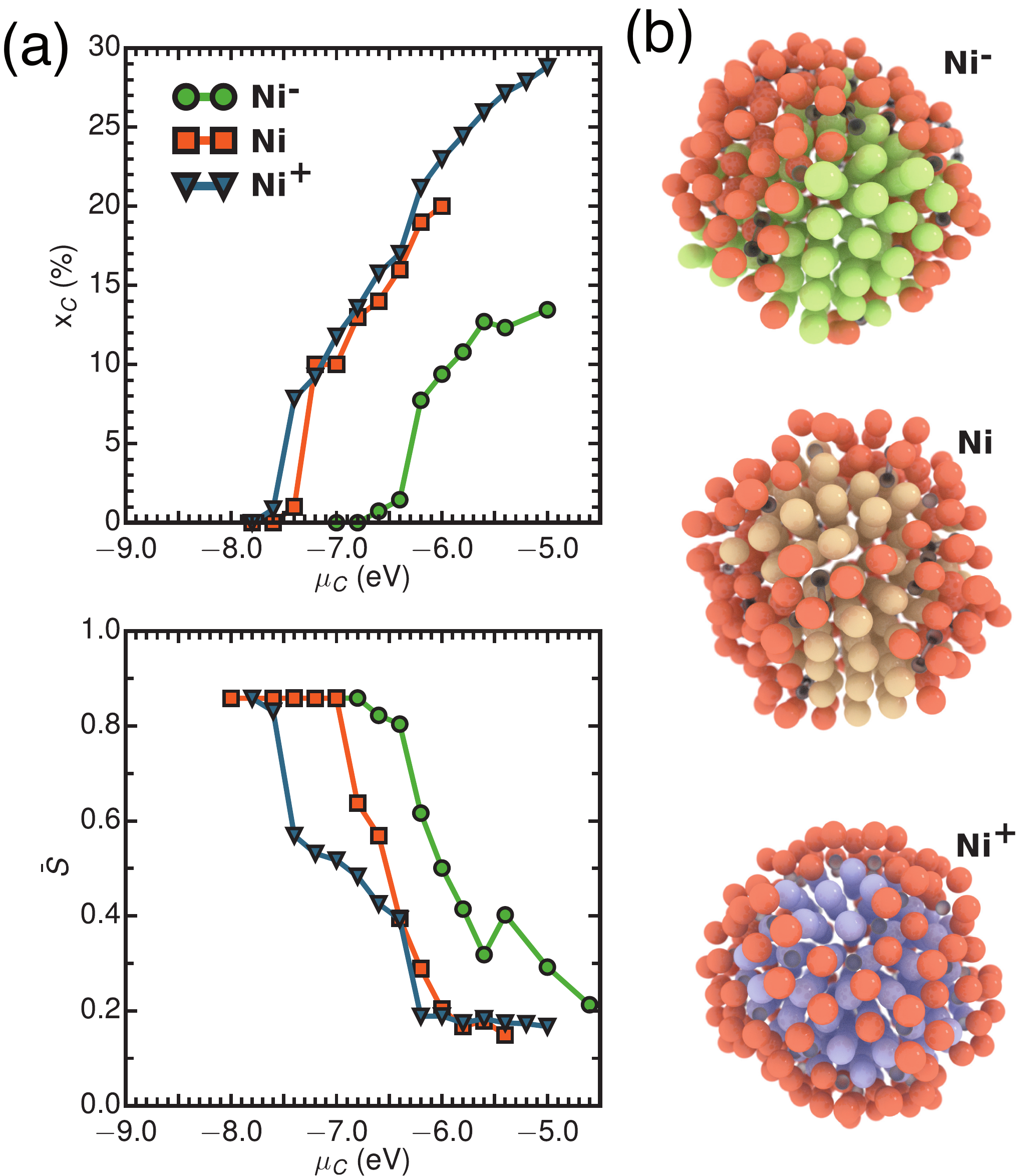}
\caption{(a) Top : C adsoprtion isotherms as a function of $\mu_c$. Bottom : average values of the order parameter as a function of $\mu_c$ for a NP initially Wulff shaped with 405 atoms. (b) Snapshot of characteristic configurations for metal-C NPs (Ni$^-$ green, Ni beige and Ni$^+$, blue atoms) surrounded by a disordered shell (orange atoms). Black balls correspond to C atoms.}
\label{Figure_3}
\end{figure}

Typical results are presented in Fig.~\ref{Figure_3}a, where the global order parameter $\bar{S}$ is plotted as a function of $\mu_{C}$ for FCC structures containing 405 atoms. $\bar{S}$ is normalized in such a way that a perfectly crystallized structure has $\bar{S}=1$, while a fully disordered one corresponds to $\bar{S}=0$. These are limiting values and we practically consider structures with $\bar{S} \ge 0.85$ as crystalline and those with $\bar{S}\le 0.35$ as amorphous. For Ni$^{+}$ NPs, $\bar{S}$ is larger than 0.7 corresponding to a mostly solid particle. Increasing $\mu_c$ induces a transition to a solid core liquid shell in between 0.3$<\bar{S}<$0.7 (see Fig.~\ref{Figure_3}b) since  C atoms induce a gradual melting of the NPs, that starts on the surface and propagates to the core. At higher $\mu_c$, a second transition is observed where the NP completely melts. In case of Ni and Ni$^{-}$, visual inspection, as well as the evolution of the order parameter as a function of $\mu_{C}$ indicate that the liquid outer layer grows continuously at the expense of the solid with increasing carbon content. By comparing with Ni$^{+}$, molten area are smaller since less C atoms can be incorporated close to the surface. 

\section{Wetting properties of nanoparticles on graphene}

From the SWNT synthesis point of view, wetting properties of the catalyst nanoparticles  are of fundamental importance to enable the nanotube growth, and, for Ni, Co and Fe catalysts, the fraction of carbon dissolved in the NP appears as a leading factor influencing them~\cite{Diarra2012}.  GCMC simulations have shown that the cap lift-off during carbon nanotube nucleation requires the weakening of the cap-catalyst interaction, 
and that sustained growth is made possible by a gradual dewetting of the NP, to avoid its encapsulation by the growing carbon $sp^{2}$  wall. Increased carbon incorporation into the catalyst was shown to facilitate these processes. The next aspect is to understand how far these wetting and interfacial properties of metallic NPs  in contact with carbon $sp^{2}$ layers can be modified by considering metals with different carbon solubilities. \\

\begin{figure}[htbp!]
\includegraphics[width=1.0\linewidth]{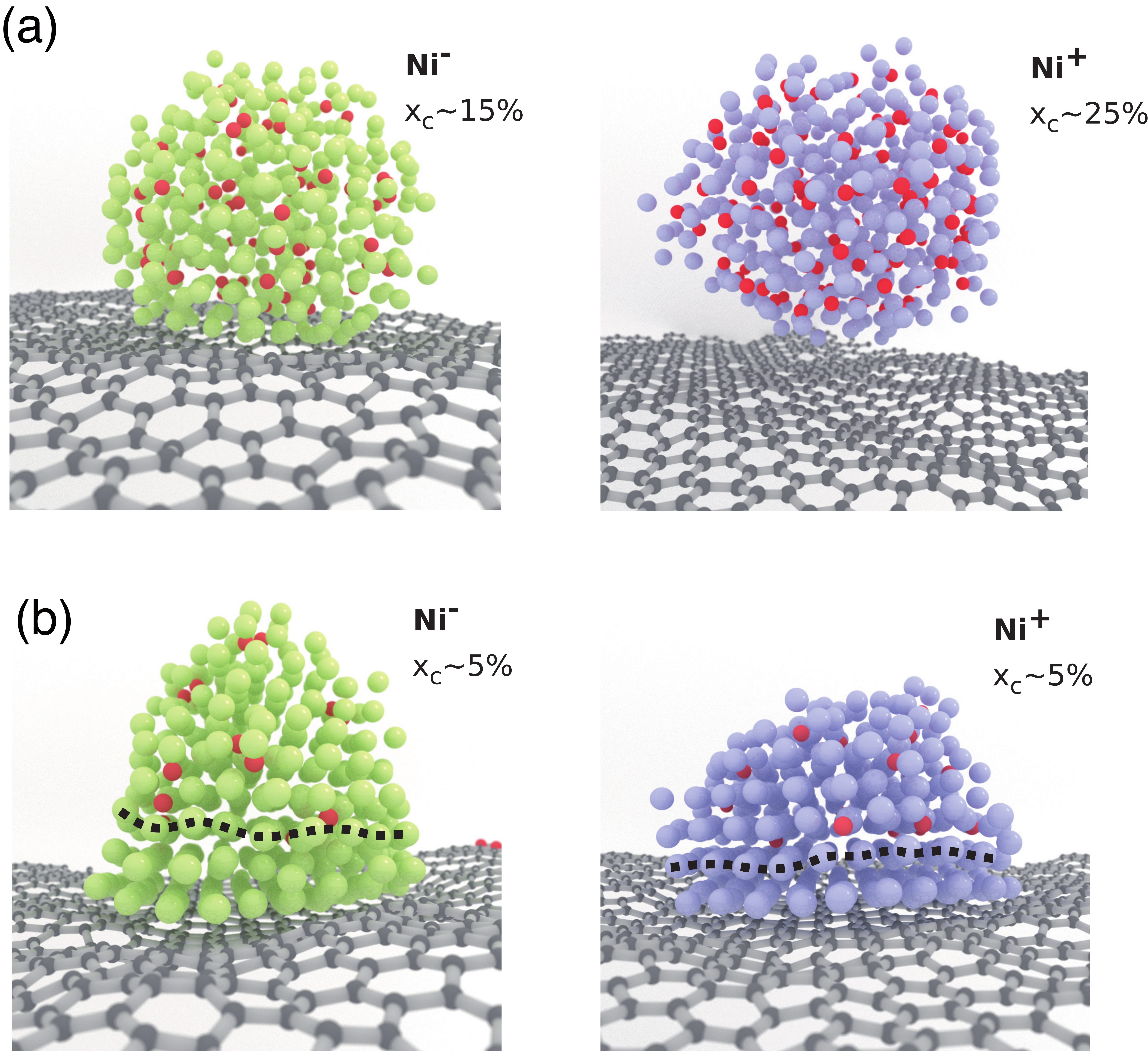}
\caption{Probing the wetting of carbon-containing metal nanoparticles deposited on graphene (a) Equilibrium structures corresponding to C-saturated NPs.  Ni$^-$ NP (left) shows a contact angle slightly larger than 90 \degree, while Ni$^+$ NP (right) is fully dewetting the substrate. (b) At intermediate concentration, we observe a carbon depletion close to the graphene layer. This absence of carbon dissolved close to the contact is more pronounced for Ni$^-$ (left), with two pure metal layers than with Ni$^+$ (right). Dashed lines delimits the C-depleted metal zone. Ni$^-$ atoms are green, Ni$^+$ atoms are blue, C atoms inside the NP are red.}
\label{Figure_4}
\end{figure}
Using canonical MC simulations at 1400 K, we investigate the wetting properties of metallic NPs with 405  atoms and C fractions from zero to the solubility limit, deposited on a graphene layer containing 1600 C atoms, using our models with three different solubilities. First, a visual inspection of the resulting structures in all cases reveals that NPs remain crystalline when no carbon atoms are dissolved inside. Contact angles for pure NPs are always smaller than 90 \degree : such NPs tend to wet the graphene layer. Focusing now on nickel (Ni) and high-carbon-solubility metal (Ni$^+$), we notice that the metallic NP tends to dewet graphene when increasing the carbon fraction dissolved in it. This is in qualitative agreement with experimental data, obtained by Naidich \textit{et al.}, showing that the contact angle of a macroscopic Ni (and also Co and Fe) drop on graphite increases with C fraction inside it~\cite{Naidich1971}. Interestingly, when this carbon fraction increases a lot, as made possible in nano-sized particles, the contact angle becomes larger until the metallic NP detaches from the graphene layer (see Fig.~\ref{Figure_4}a). In case of Ni$^{-}$ NPs, the contact angle remains close to $\sim90\degree$ whatever the carbon concentration, meaning that no detachment is observed. Their inabilities to dewet is due to the limited carbon content in such NPs ($\sim15 \%$ C), as highlighted by the isotherms presented in Fig.~\ref{Figure_2}. \\

In addition to the wetting/dewetting tendency, we notice that dissolved carbon atoms are not homogeneously distributed within the NPs (see Fig.~\ref{Figure_4}b). Their distributions are notably altered, with the proportion of C atoms in subsurface sites significantly reduced, compared to the distribution throughout the rest of the NP. As seen in Fig.~\ref{Figure_4}b, no C atoms are present close to the interface with graphene, while carbon remains distributed throughout the rest of the catalyst. This means that dissolved C atoms tend to avoid the graphene layer,  while it is well known that subsurface interstitial sites are preferred locations for individual C atoms below a free surface~\cite{Amara2006}. Such a depletion effect was also observed during graphene growth on Ni (111)~\cite{Weatherup2014, Benayad2013}. By combining MC simulations, DFT-based calculations and \textit{in-situ} X-ray photoelectron spectroscopy, we could evidence this depletion effect and revealed an interdependency between the carbon distribution close to the catalyst surface and the strength of the graphene-Ni interaction. Epitaxial graphene formation on Ni(111) leads to a depletion of carbon close to the Ni surface~\cite{Weatherup2014}. 

In the present situation, this effect leads locally to the interaction of graphene with pure NP which depends on the metal-carbon interaction. To test this for our three metal-carbon models, we calculate the binding energy of graphene on a (111) pure metallic surface. Since the lattice parameters of Ni$^-$ and Ni$^+$ are the same as that of Ni, the (111) surfaces of these metals are in almost perfect epitaxy with graphene. For our three models, very small differences (less than 2\%) are observed. Adhesion energies remain close to -0.32 eV per carbon atom, a bit larger than the experimental one for Ni~\cite{Batzill2012}. Such values favor the adhesion of the NP on graphene, and a dewetting tendency can only result from the presence of subsurface carbon close to the graphene layer. 

\begin{figure}[htbp!]
\includegraphics[width=0.80\linewidth]{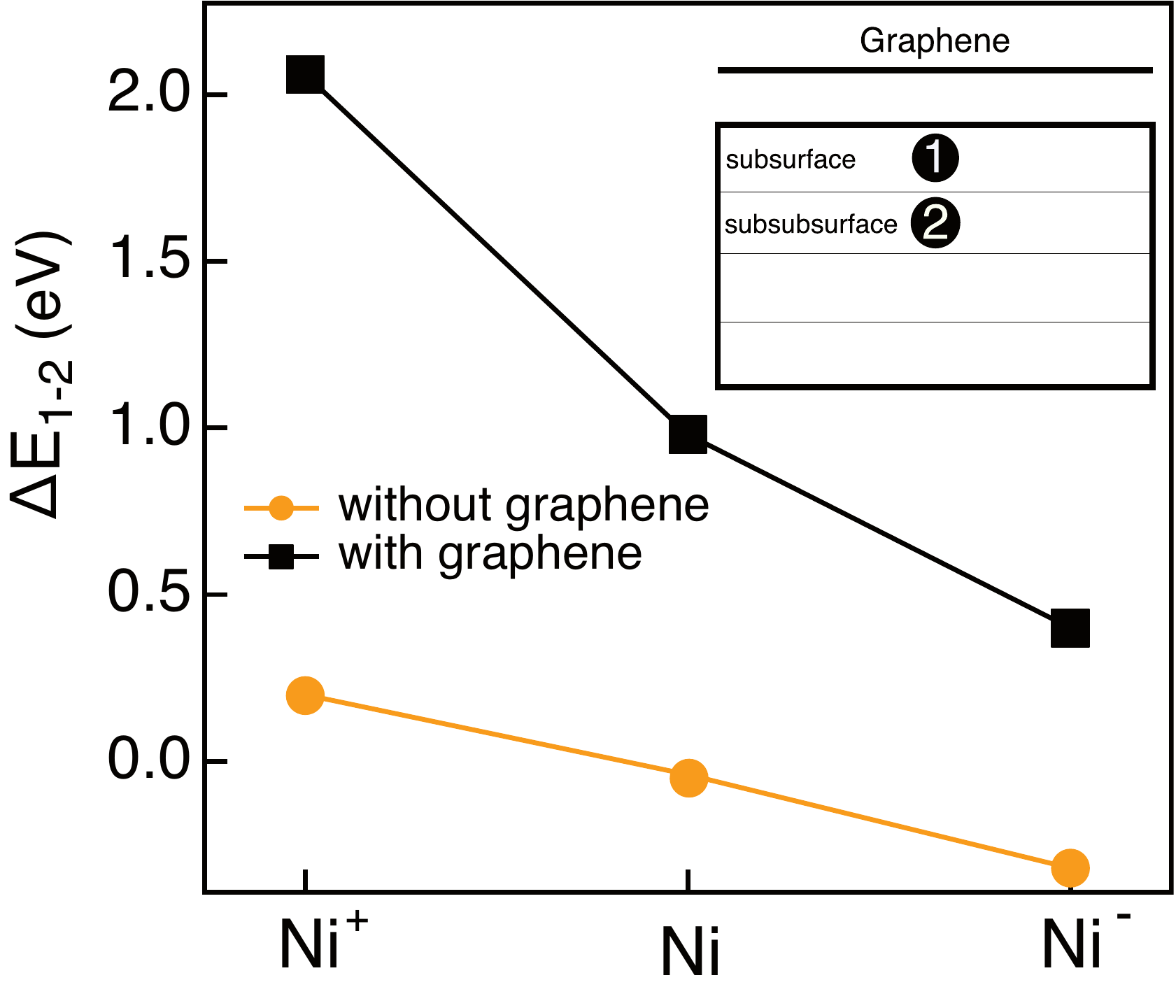}
\caption{Differences in carbon dissolution energies ($\bigtriangleup E_{1-2}$), for a carbon atom located in subsurface (1) or sub-subsurface (2) interstitial site, without (yellow) and with (black) graphene covering a (111) slab of "nickel" with different C solubilities. Inset: Sketch of the epitaxial graphene covered (111) slab, with the interstitial sites indicated.}
\label{Figure_5}
\end{figure}

Comparing our three models, we notice that the depletion of carbon close to  graphene is more pronounced for Ni$^{-}$, where two layers of metal are carbon-free, as shown in Fig.~\ref{Figure_4}b. To quantify it, energy differences between an interstitial C atom occupying a subsurface and a sub-subsurface site ($\bigtriangleup E_{1-2}$) in the (111) slab are calculated by performing simulated annealing at 0 K. To make the discussion easier, we define subsurface (1) and sub-subsurface (2) interstitial sites as those located respectively between the first and second, and the second and third planes from the surface, both providing a full octahedral environment. Results are presented in Fig.~\ref{Figure_5}. Without graphene coverage, $\bigtriangleup E_{1-2}$ remains close to zero. For Ni and Ni$^{-}$ subsurface C site is most stable than the sub-subsurface one, while the opposite stands for Ni$^{+}$. With an epitaxial graphene cover on the metallic surface, sub-subsurface position is always preferred, as in our calculations and experiments on Ni. More interestingly, in case of Ni$^{+}$, the difference between both situations is really significant ($\sim2$ eV) explaining that the presence of a graphene layer below the NP induces an important depletion of C dissolved close to the surface. We thus understand that the adhesion of the NP on graphene is directly correlated to the distribution of C atoms close to the surface, a carbon depletion promoting stronger adhesion and wetting. 

\section{Application : growth of carbon nanotubes}

During	 CVD, nanotubes grow from a NP that is either supported by a substrate or in contact with the tube only (tip growth or floating catalyst reactor). The decomposition of  carbon-bearing precursors takes place at high temperatures (500-1200$\degree$C) in a wide pressure range, depending on the precursor's reactivity. This involves complex thermochemical reactions that cannot be correctly described in an full atomistic computer simulation of  SWNT growth. However, when focusing on thermodynamic aspects of growth, the important point is that these reactions lead to the presence of carbon atoms, close to the surface of the catalyst, at a given chemical potential~\cite{Snoeck1997}. \\
\begin{figure}[htbp!]
\includegraphics[width=0.80\linewidth]{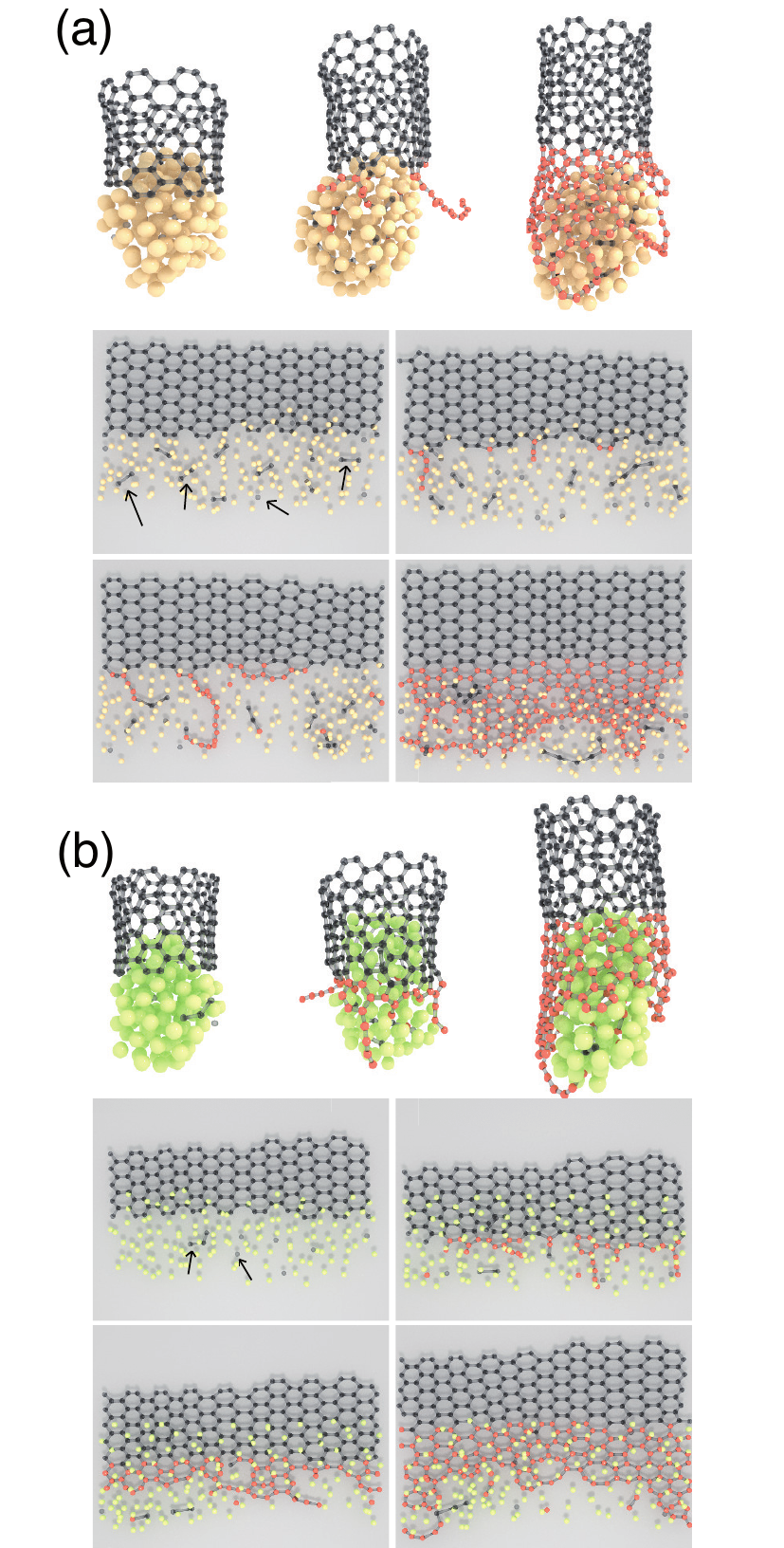}
\caption{Snapshots of atomic configurations during the growth of a tube butt on a NP (C atoms are in black, added C atoms are in orange and some C atoms in subsurface positions are indicated with arrows). Arrows indicate the presence of C atoms in subsurface position. (a) During SWNT growth from Ni (beige) NP, longer chains, bound on one end and prone to float away from the surface are observed. This leads to more defective $sp^2$ structures. (b) From Ni$^{-}$ nanoparticle, smaller chains remain more strongly bound to the NP surface. Because of this, defect-free $sp^2$ network is more easily formed.}
\label{Figure_6}
\end{figure}
In order to see how SWNT growth mechanisms are influenced by  modifications of carbon content inside the NP, we chose to compare Ni and Ni$^{-}$ catalysts. We start by fixing short nanotube butts with 173 C atoms on pure NPs with 120 metal atoms. Tubes are chosen with diameters in the 1.2 nm range and different chiralities. We used them as starting configurations for a series of GCMC simulations at different temperatures with $\mu_{C}$ conditions chosen to trigger tube growth. To get a better insight into the growth mechanisms, MC simulations are visualized by unrolling the systems along the tube axis~\cite{Yoshikawa2016}. In this way,  microscopic mechanisms at the atomic scale can be revealed. 

Fig.~\ref{Figure_6} presents a series of atomic configurations leading to successful growth on a Ni nanoparticle. During the simulations, C atoms diffuse on the surface to form chains that can crawl on, or close to the surface and cross each other. At their intersections, threefold coordinated C atoms act as nucleation centers for C $sp^{2}$ structures to develop. Since the surface layers of the NPs are saturated with carbon, these $sp$ carbon chains interact weakly with the underlying surface. As a result, chains tend to detach from the surface, like a garland, leading to the formation of somewhat defective structures, as seen in Fig.~\ref{Figure_6}a. In addition, the NP tends to escape from the tube, because of the dewetting tendency induced by the carbon dissolved in the NP, while the growing walls tend to catch it back. This sometimes leads to a complete detachment for some particular ($\mu_{C}$, T) conditions. 

Let us now consider NPs with a reduced C solubility (Ni$^{-}$). In several cases, we notice the formation of tubes containing few defects (see Fig.~\ref{Figure_6}b). Remarkably, the growth of three rows of defect-free hexagonal rings has been obtained, corresponding to a 1.2 nm extension of the SWNT. Here, incoming C atoms form dimers and short chains in stronger interaction with the surface of the Ni$^{-}$ NP. Because they are more strongly bound to the surface, these short chains quickly form cycles, pentagons first, that readily transform into hexagons, leading to graphene-like walls with less defects. Interestingly, we have identified subsurface-carbon depleted zones, as particular areas where defect-free graphene is formed (see Fig.~\ref{Figure_6}b).  Contrary to the previous case, unsuccessful growth attempts are often caused by a too strong adhesion of the growing wall on the NP,  leading to particle encapsulation.

Successfully growing defectless tube sections is important to discuss chirality control at the atomic scale.  Until now, whether the method employed to study nucleation-growth mechanism of SWNT is empirical~\cite{Shibuta2003,Ding2004, Neyts2011,Khalilov2015} or semi-empirical~\cite{Amara2008, Page2015},  final configurations are plagued by a high concentration of atomic-scale defects, i.e., nonhexagonal rings, adatoms, or vacancies. We show here that fine tuning ($\mu_{C}$, T) conditions, adapted to the catalyst used, can lead to highly crystalline tube structures that make it possible to analyze the atomistic growth mechanisms and their possible connection to chiral selectivity.

\section{Discussion and Conclusion}

\begin{figure}[htbp!]
\includegraphics[width=1.00\linewidth]{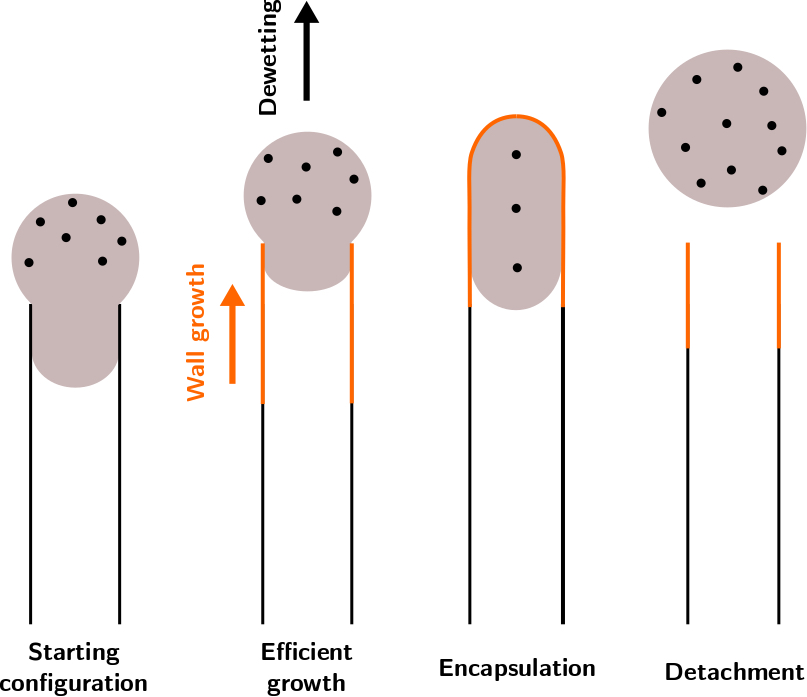}
\caption{Sketch of SWNT growth. Efficient growth is observed when wall growth and NP dewetting proceed in a sustainable way. Particles containing a low C fraction can be easily encapsulated, thus favoring growth deactivation. On the contrary, carbon saturated NPs interact less strongly with the growing tube and are thus more prone to detachment.}
\label{Figure_7}
\end{figure}

Sustainable growth conditions, and two possible scenarios for growth termination discussed above are sketched in Fig.~\ref{Figure_7}. Two competing phenomena can be identified, namely NP dewetting and wall growth. Thermodynamic analysis enables to identify the correct driving force to incorporate C atoms in the carbon $sp^2$ wall, and in the catalyst, thus modifying its interfacial properties. In the case of Ni-like catalysts discussed above, a proper choice of a catalyst with the right carbon solubility, that would be in between Ni and Ni$^{-}$ in the present instance, would lead to a right balance between wall growth and NP dewetting, and hence sustainable growth. In real experiments, carbon solubility could be tuned using bimetallic alloys, forming a real nanoalloy, as opposed to phase separated NPs, such as Ni$_x$Fe$_{(1-x)}$~\cite{Chiang2009}. Another way to act on it might be the use of a surfactant, such as sulfur, that can be introduced in the system with the gaseous phase, via thiophene molecules, as done by Sundaram \emph{et al.}~\cite{Sundaram2011}.

For low carbon fractions, while carbon wall dewetting is a necessary step for SWNT nucleation and prolongation, the strong adhesion energy favors the adhesion and wetting of the carbon $sp^{2}$ wall on the surface of the nanoparticle, thus causing the catalyst particle deactivation. In contrast, high-carbon-solubility metal nanoparticles have a weak adhesion with tube wall, resulting in the detachment of the catalyst from the tube that causes the growth to stop. Between these two extreme situations, efficient growth can be observed when both NP dewetting and wall growth smoothly proceed together in a sustainable way. Beyond the optimization of CVD process parameteres, such as temperature and feedstock pressure, our present calculations suggest that stable growth conditions can also be searched for by  tuning the carbon solubility in the catalyst NP.
Recently, we have confirmed this mechanism by demonstrating that low carbon fractions in the catalyst, resulting from the use of CH$_4$ feedstock, lead to tangential growth, where the metallic nanoparticle wets the tube inner wall and can be easily encapsulated by the graphitic layers, facilitating the deactivation of the catalyst and the termination of SWNT growth~\cite{He2017}. In such a situations, shorter tubes were found to be formed. On the contrary, using CO as a feedstock, catalyst nanoparticles with a high carbon fraction nucleate SWNTs by perpendicular mode and can keep their activity for much longer time, accounting for the formation of long SWCNTs.  This offers new possibilities to tune the catalytic activities of metal NPs by improving its abilities to dissolve carbon and contributes to a better understanding of SWNT growth mechanisms, paving the way to a rational search for better catalysts for controlled synthesis of SWNTs.

\section*{Acknowledgments}
The research leading to these results has received funding from the European Union Seventh Framework Programme (FP7/2007-2013) under grant agreement n\textsuperscript{o} 604472 (IRENA project) and French Research Funding Agency under grant No. ANR-13-BS10-0015-01 (SYNAPSE project). Dr F. Ducastelle is acknowledged for many fruitful discussions.

\end{document}